\documentclass[epj]{svjour}
\usepackage{latexsym}
\usepackage{graphics}
\begin{document}
\title{Surface Plasmons on a Quasiperiodic Grating}

\author{J. Milton Pereira Jr., G. A. Farias, and R. N. Costa Filho}

\institute{Departamento de F\'{\i}sica, Universidade Federal do
Cear\'a, Caixa Postal 6030\\ Campus do Pici, $60451-970$
Fortaleza, Cear\'a, Brazil}

\date{Received: date / Revised version: date}

\abstract{
A method is presented for calculating the frequencies of
non-retarded surface plasmons propagating on
a semi-inifinite medium with a surface profile described by a one-dimension
quasiperiodic function.
The profiles are generated, in analogy with previous work on quasiperiodic
superlattices, by repeating unitary cells constructed according to an inflation rule.
Dispersion relations are obtained for a semi-infinite free-electron metal
as the active medium, with surface profiles obeying the Fibonacci and Thue-Morse
sequences.
}
\maketitle

\section{Introduction}
\label{intro}
Extensive experimental and theoretical work on
quasiperiodic
systems has revealed interesting new properties which are
not found in
either periodic or in disordered systems. Examples of quasiperiodic
systems are artificial structures such as surfaces with quasiperiodic
tilling patterns \cite{Torres}, as well as multilayers constructed by
juxtaposing different building blocks following quasiperiodic
inflation rules \cite{Ela+Mike}. These structures are generally
considered as intermediate states between ordered and disordered
systems. 
It has been found that
the energy spectra of elementary excitations (e.g. polaritons,
phonons, spin waves) of such structures
are highly fragmented, with a self-similar
pattern (for a comprehensive review, see Ref.\cite{Ela+Mike} and
references therein).
In fact, this property is now accepted as a signature of a
quasiperiodic
system.

Another topic that has attracted considerable attention is the study of the
propagation of surface modes on disordered media.
Among these excitations, surface plasmons, which are electronic
collective modes bound to a dielectric-metal interface, have been
some of
the most intensively studied.
Due to their surface nature, these modes are strongly
influenced by
the shape of the interface along which they propagate.
In fact, the
excitation of surface plasmon modes on randomly rough surfaces can lead to
localization effects \cite{Mara2}, and is also associated
with surface phenomena such as the enhanced
second harmonic generation \cite{Bozh1}.
Recent studies \cite{Mara1,Bozh} of
surface plasmons on periodically corrugated surfaces have indicated the
presence of absolute band gaps in the frequency spectrum, which in turn show
a dependence on the geometry of the surface.

The aim of this work is to present a method for calculating
the spectrum of excitations of
a metal-dielectric
interface with a quasiperiodic grid of parallel
ridges. In this case, since the quasiperiodic elements are restricted to
the surface of the medium, they
may exert a significant influence on its spectrum
of surface excitations.
Specifically, we calculate the surface plasmon spectrum of surfaces with
one-dimensional profiles described by functions
created by substitutional sequences.
We use these functions to construct a series of periodic profiles,
with each consisting of a unitary cell obtained from a quasiperiodic sequence, repeated
along the $x_1$ direction. For each generation of
the sequence, the period length $a$ increases, and the
quasiperiodic case is
obtained in the limit $a \rightarrow \infty$.

\section{Model}
\label{sec:1}
The theory is based on the Rayleigh hypothesis formalism described
in Ref. \cite{Glass}, which has been shown to give exact results for surfaces
with small corrugations (i.e. with height to width ratios $< 0.072$)
defined by analytic functions.
In this formalism, solutions of Laplace's equation are found that
vanish as the distance ($|x_3|$) to the interface increases.
For a semi-infinite medium with a surface described by a one-dimensional
periodic profile, the potentials associated with modes propagating
along the $x_1$ direction can be assumed to obey the Bloch property
\begin{equation}
\phi(x_1+a,\omega) =  \phi(x_1,\omega) e^{ika},
\end{equation}
where $a$ is the period of the surface.
Thus, the potential associated with a surface mode can be
expressed as,
\begin{equation}
\phi(x_1,x_3)=\sum_n^\infty C_n \exp(ik_nx_1+\alpha_nx_3)
\end{equation}
where $\alpha_n = |k+2\pi n/a|$, $k_n = k+2\pi n/a$, $k$ is the
wavevector, $x_3$ is the direction normal to the surface of the
active medium and the $C_n$ factors are the Fourier amplitudes of
the potential. By applying the interface boundary
conditions, the frequencies of the non-retarded
surface plasmons can be obtained by solving the eigenvalue equation
\begin{equation}
\sum_{p=-\infty}^\infty \, M_{rp}(k)C_p = {{\epsilon (\omega)+1}\over
{\epsilon (\omega)+1}}C_r ,
\end{equation}
with the elements of the matrix $M_{rp}(k)$ being given by
\begin{eqnarray}
M_{rp}(k)={{\alpha_r(k)\alpha_p(k)-k_rk_p}\over
{\alpha_r(\alpha_r-\alpha_p)}}K(\alpha_r-\alpha_p,k_r-k_p)&&\cr
&&
\end{eqnarray}
for $r \neq p$ and $M_{rp}(k)= 0$ otherwise.
The function
$K(\alpha_r-\alpha_p,k_r-k_p)$ is obtained from the integral
\begin{equation}
K(\alpha,k_l) = {1\over a}\, \int_{-a/2}^{a/2} e^{\alpha \zeta (x)}e^{-ik_lx}dx ,
\end{equation}
where we used $\alpha \equiv \alpha_r-\alpha_p$ and $k_l\equiv k_r-k_p$.
The function $\zeta (x)$ describes the surface profile.

In the following calculations, we employ a surface model designed in a similar
fashion to the structures used in the multilayer calculations.
The goal is to construct a profile that approximates a quasiperiodic surface
by using a known function, taken in finite intervals, in order to define
a {\it periodic} function. Thus a series of periodic surfaces can be
described, each consisting of unitary cells made of building blocks
arranged in a finite quasiperiodic sequence, repeated along the $x_1$
direction. The overall periodic nature of the surface allows us to use the formalism
of Glass {\it et al.}. However,
for each successive generation of the sequence, the period
length $a$ increases, and an actual quasiperiodic
surface is obtained as the length of the unitary cell
grows to infinity.
In the present case, the building blocks are chosen to be
sinusoidal functions. These functions are defined in intervals such
that each block corresponds to a ridge on the surface of the
metal. This approach to the definition of the profile is similar to
the one used in Ref \cite{EPJ} for isolated ridges on a flat surface.
The ridges can be of two types, with different heights or
widths, which we henceforth refer to as blocks 'A' and 'B' (see Fig. 1).
These sinusoidal blocks are described by the functions
\begin{equation}
\zeta_a(x_1)=2A_a \cos^2 (\pi x_1/L_a)
\end{equation}
in intervals $s_aL_a+s_bL_b-L_a/2 < x < s_aL_a+s_bL_b+L_a/2$,
for 'A' ridges, and
\begin{equation}
\zeta_b(x_1)=2A_b \cos^2 (\pi x_1/L_b),
\end{equation}
in intervals $s_aL_a+s_bL_b-L_b/2 < x < s_aL_a+s_bL_b+L_b/2$, for the 'B' ridges,
with $s_{a,b}=0,1,2,...$. In order
to construct the unitary cell, one has only to specify the intervals
corresponding to each block.

\begin{figure}
\centering{\resizebox*{!}{5.5cm}{\includegraphics{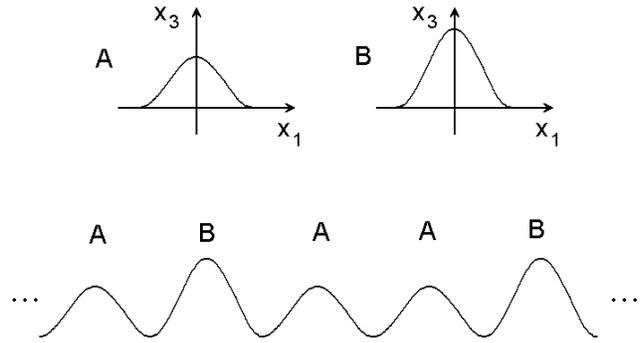}}}
\caption{Schematic representation of the surface profile used. Ridges, labeled
$A$ and $B$ are arranged periodically on a flat surface. The unitary cell is
constructed according to an inflation rule. The lower image represents a unitary
cell corresponding to the 5th term in the Fibonacci sequence.} \label{fig:wn}
\end{figure}

By following the prescriptions above, one can then obtain
the kernel in Eq.(3), by applying suitable changes of variables in Eq.(5), as
\begin{equation}
K(\alpha,k_l) = f_a(k_l)K_a(\alpha,k_l)+f_b(k_l)K_b(\alpha,k_l) ,
\end{equation}
where
\begin{equation}
K_{c}(\alpha,k_l) = {1\over a}\, \sum_{n=0}^\infty
{\alpha^n \over {n!}}\int_{-L_{c}/2}^{L_{c}/2} \zeta_{c}^n (x_1)e^{-ik_lx_1}dx_1 ,
\end{equation}
with $c=a,b$. The functions $f_a(k_l)$ and $f_b(k_l)$ contain
the information concerning the positions of the ridges in the
unitary cell, and therefore depend on the sequence generation.
In this paper we consider the case where
$A_a \neq A_b$ and $L_a = L_b$.

Let us first consider the Fibonacci sequence. In this case,
we assume $S_0 = B$, $S_1 = A$ as the first two terms in the sequence.
The other terms are generated by an operation that consists of juxtaposing
the two previous terms in the sequence, following the rule
\begin{equation}
S_{n+1} =S_n\,S_{n-1}.
\end{equation}
Thus, the next three terms are $S_2 = AB$, $S_3 = ABA$,
$S_4 = ABAAB$ etc. Therefore, for the $S_3$ generation we obtain
an unitary cell of length $a = 3L$. This cell can be
defined as
\begin{equation}
\zeta(x_1)=\left\{
\begin{array}{ccc}
\zeta_a(x_1), && -3L/2 < x < -L/2, \\
&&\\
\zeta_b(x_1), && -L/2 < x < L/2, \\
&&\\
\zeta_a(x_1), && L/2 < x < 3L/2
\end{array}
\right.
\end{equation}

In this particular case, there is a B ridge centered at the origin, a A ridge shifted
$L$ units to the left of the origin, and another A ridge shifted by $L$
units to the right. That translates as
\begin{equation}
f_b(k_l) = 1 \qquad , \qquad  f_a(k_l)=2\cos(k_lL).
\end{equation}
For the sinusoidal profiles considered here, it can be shown that,
\begin{eqnarray}
&&K_{c}(\alpha,k_l)=\sum_{n=0}^\infty {{\xi^n}\over {n!}}
\Bigl({{A_c}\over {L}}\Bigr)^nF^{-n}\times\cr&&\sum_{m=0}^{2n}
\Bigl({2n \atop m}\Bigr){
{\sin[\pi(n-m)-\pi(r-p)/F]}\over {\pi[F(n-m)-(r-p)]}},
\end{eqnarray}
where $F$ is the total number of ridges in the unitary cell.

This formalism can be extended to profiles generated by other
sequences in a straightforward way. In the case of the Thue-Morse sequence,
one applies the following substitution rules:
\begin{equation}
S_{n} =S_{n-1}\,S_{n-1}^+
\end{equation}
and
\begin{equation}
S_{n}^+ =S_{n-1}^+\,S_{n-1}
\end{equation}
for $n\geq 1$, with $S_0 = A$ and $S_0^+=B$. Thus, the next three generations
of this sequence are: $S_1=AB$, $S_2=ABBA$ and $S_3=ABBABAAB$. Thus, the number
of blocks in a unitary cell corresponding to a sequence generation $S_n$ is $2^n$.

\section{Numerical Results}
\label{sec:2}

We now present numerical results for gratings ruled on a free-electron metal, with
the dielectric constant
\begin{equation}
\epsilon(\omega) = 1 -
\frac{\omega_p^2}{\omega^2},
\end{equation}
where $\omega_p$ is the plasma frequency of the
conduction electrons in the metal. The surface plasmon spectra for sinusoidal gratings
are known to display an infinite set of dispersion branches distributed symmetrically
around the flat-surface-plasmons frequency $\omega_{fsp}=\omega_p/\sqrt{2}$ \cite{Glass}.
This behaviour is also found in the present results.
The graphs in Fig. 2 show a comparison of the dispersion relations of surface plasmons
for gratings constructed with the Fibonacci sequence, for increasing generation numbers,
using $A_a/L = 0.04$ and $A_b/L = 0.07$. Only higher frequency
branches are shown. The frequencies are normalised by $\omega_{fsp}$
and the wavevectors are normalised by $2\pi / a$. As the number of
blocks in the unitary cells increases, the convergence of the results
becomes slower, and more terms must be added to the summation in Eq.
(3). However, even for S$_8$, a good convergence can be obtained for
$N < 100$ terms. Graph (a) shows the periodic case, which corresponds
to a sinusoidal grating with $A/L = 0.07$. Graphs (b) to (d) show the
results for sequence generations S$_4$ ($F=5$), S$_6$ ($F=13$) and
S$_8$ ($F=34$), respectively. In contrast with the periodic case, in
which most of the branches are found in close proximity to
$\omega_{fsp}$, as the generation number increases, several
high-frequency branches are found and one can observe a tendency to
the formation of frequency bands for large generation numbers. The
results also indicate the presence of several frequency gaps arising
in the plasmon spectrum. This result is consistent with what has been
previously observed for excitations in multilayer systems.
Nevertheless, in the present case the width of the gaps is directly
related to the height to width ratios of the different blocks, a
parameter that can, in principle, be varied continuously, whereas in
multilayer systems the gaps depend on the different properties of each
block, which cannot be as easily controlled.

Two results for gratings following the Fibonacci sequence are displayed in Fig. 3.
Both graphs show some of the high-frequency dispersion branches of surface plasmons on
gratings with unitary cells corresponding to the $S_9$ term in the sequence ($F = 34$). However, for
the graph labeled (a), the values $A_a/L = 0.04, A_b/L = 0.07$ were used, whereas for graph
(b) the parameters were $A_a/L = 0.07, A_b/L = 0.04$. One can see from the graphs that,
in contrast with the three high-frequency bands that arise in the $A_a > A_b$ case, the
results for $A_a < A_b$ show the appearance of a single high-frequency band.
These results illustrate the inherent asymetry of the Fibonacci sequence, which is
a consequence of the fact that, in that sequence, the number of $B$ blocks is found to be
always smaller than the number of $A$ blocks (for large generation numbers, the ratio of the number
of $B$ blocks and the number $A$ blocks approaches the number $\phi \approx 0.618$).
In addition, in the Fibonacci sequence, the $B$ blocks are always located between $A$
blocks. This asymmetry does not occur in
the Thue-Morse sequence, in which the number of $A$ and $B$ blocks is always the same.

Figure 4 shows the upper dispersion branches for surface plasmons propagating on a
grating with a unitary cell corresponding to the 5th generation of the Thue-Morse
sequence, for $A_a/L = 0.02$ and $A_b/L = 0.07$. As in the previous cases, the graph
shows several gaps in the spectrum, with three groups of high-frequency branches
separated by large gaps, with several small gaps located in the high-frequency regions.
This behavior points to a fractal nature of the spectrum as the generation number
increases.

\begin{figure}
\centering{\resizebox*{!}{8.5cm}{\includegraphics{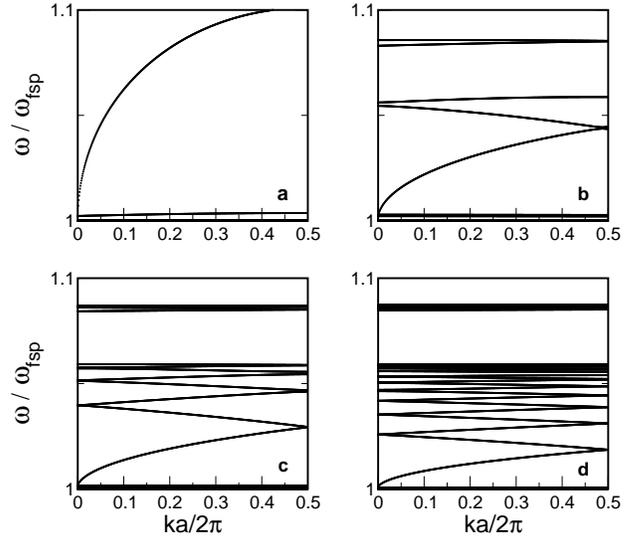}}}
\caption{Surface plasmon frequencies as a function of wavevector for
a sinusoidal grating (a), with $A/L = 0.07$,
and gratings with unitary cells obtained from
the Fibonacci rule: (b) S$_4$, (c) S$_6$ and (d) S$_9$, for $A_a/L = 0.04$
and $A_b/L = 0.07$.} \label{fig:wn}
\end{figure}

\begin{figure}
\centering{\resizebox*{!}{8.5cm}{\includegraphics{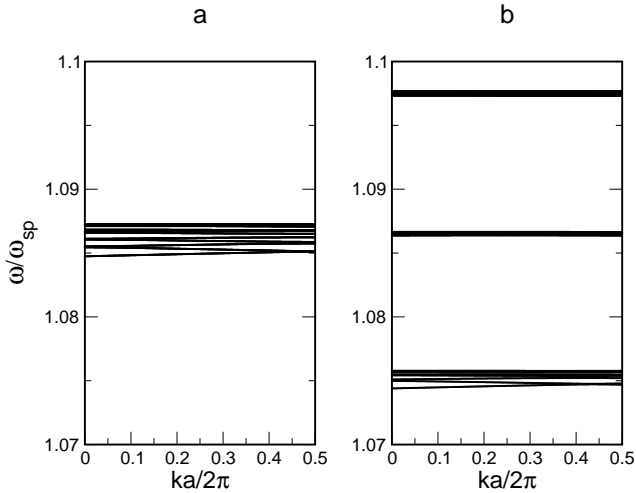}}}
\caption{High-frequency surface plasmon dispersion branches for
gratings with a unitary cell corresponding to the 9th generation of the
Fibonacci sequence, for (a)
$A_a/L = 0.04$, $A_b/L = 0.07$ and (b) $A_a/L = 0.07$, $A_b/L = 0.04$.} \label{fig:wn}
\end{figure}

\begin{figure}
\centering{\resizebox*{!}{8.5cm}{\includegraphics{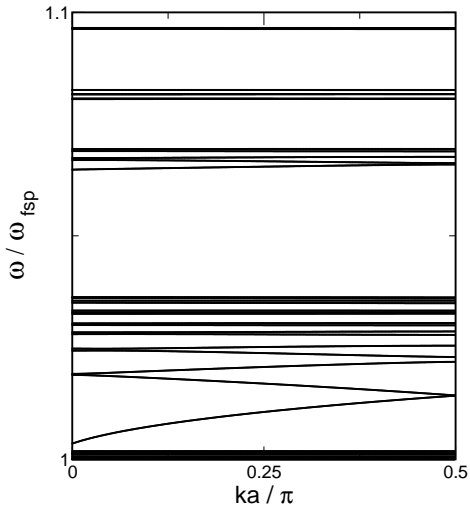}}}
\caption{Surface plasmon dispersion branches for a grating with a unitary
cell obtained from the Thue-Morse sequence (S$_5$), for
$A_a/L = 0.02$, $A_b/L = 0.07$.} \label{fig:wn}
\end{figure}

\section{Conclusions}

In summary, we calculated the frequencies of surface plasmons
propagating on gratings with one-dimensional surface profiles that
follow quasiperiodic sequences.
The surface morphology is described by a model
that allows us to employ methods previously developed
for the study of surfaces with ordered profiles, and at the same time to explore
a topography that displays deterministic disorder.
The results were obtained for gratings constructed with Fibonacci and
Thue-Morse sequences and can be easily extended to other sequences.
The surface plasmon dispersions show a large number of gaps
that are related to the sequence generation and to the
aspect ratio of the ridges that form the grating.
The spectra have
a fractal character that is similar to what has been previously observed
for other excitations in quasiperiodic structures.
The method presented here allows the calculation of the frequencies of {\it propagating}
modes. However, one
cannot rule out the possibility of existence of {\it non-propagating} plasmon modes
in such structures.
In fact, such localized modes are often found in
quasiperiodic superlattice structures, with frequencies located in the gaps of the spectrum.
This method can also be applied to the study of other surface
excitations, such as surface polaritons, and to investigate
the light scattering properties of quasiperiodic surfaces.
Surfaces with periodic features have been constructed by photolithography,
and such a technique could in principle be applied in order to create the
quasiperiodic structures
described here.

The authors would like to acknowledge the financial support of the
brazilian council for research (CNPq).

\end{document}